\documentclass{llncs}
\usepackage{times}
\usepackage{graphicx}
\usepackage{epsfig}
\usepackage{subfigure}
\usepackage{latexsym}
\usepackage{paralist}

\newcommand{\fullver}[1]{}
\newcommand{\ex}{{\rm E}}

\renewcommand*{\Pr}{\mathop{\mathrm{Pr}}}

  \newcommand{\goesto}{\rightarrow}

\def\pminus{\pm}

\def\AND{\wedge}
\def\OR{\vee}

\def\goesto{\rightarrow}
\def\implies{\Rightarrow}
\def\tcimplies{\stackrel{*}{\rightarrow}}

\def\beginproof{\noindent{\bf Proof.}\quad}
\def\endproof{}

\def\qed{\hfill$\Box$\newline\vspace{5mm}}

\begin{document}

\title{Satisfying Assignments of Random Boolean CSP: Clusters and Overlaps}
\author{Gabriel Istrate}
  \institute{ eAustria Research Institute, Bd. Corneliu Coposu 4, \\
  cam. 045B, Timi\c{s}oara, RO-300223, Romania \\
  \email gabrielistrate@acm.org \\}
  \vspace{-5mm}
 \maketitle \vspace{-5mm}
\begin{abstract}
{\small The distribution of overlaps of solutions of a random CSP is
an indicator of the overall geometry of its solution space. For
random $k$-SAT, nonrigorous methods from Statistical Physics support
the validity of the ``one step replica symmetry breaking'' approach.
Some of these predictions were rigorously confirmed in
\cite{cond-mat/0504070/prl} \cite{cond-mat/0506053}. There it is
proved that the overlap distribution of random $k$-SAT, $k\geq 9$,
has discontinuous support. Furthermore, Achlioptas and
Ricci-Tersenghi \cite{achlioptas-frozen} proved that, for random
$k$-SAT, $k\geq 8$. and constraint densities close enough to the
phase transition:
\begin{itemize}
 \item there exists an exponential number of clusters of satisfying assignments.
 \item the distance between satisfying assignments in different clusters is linear.
\end{itemize}

We aim to understand the structural properties of random CSP that
lead to solution clustering.  To this end, we prove two results on
the cluster structure of solutions for binary CSP  under the random
model from \cite{molloy-stoc2002}:

\begin{enumerate}
\item For all constraint sets  $S$ (described in \cite{creignou-daude-threshold,istrate-sharp}) s.t. $SAT(S)$ has a
sharp threshold and all $q\in (0,1]$, $q$-overlap-$SAT(S)$ has a
sharp threshold (i.e. the first step of the approach in
\cite{cond-mat/0504070/prl} works in all nontrivial cases).
\item For any constraint density value $c<1$, the set of solutions of a
random instance of 2-SAT form, w.h.p., a single cluster. Also, for and any $q\in (0,1]$
 such an instance has w.h.p. two satisfying assignment of overlap $\sim q$. Thus, as
expected from Statistical Physics predictions, the second step of
the approach in \cite{cond-mat/0504070/prl} fails for
2-SAT.
\end{enumerate}}
\end{abstract}
\vspace{-5mm}
\section{Introduction}

\begin{figure}

\begin{center}
\begin{minipage}[l]{6cm}
\includegraphics[height=4cm,width=6cm]{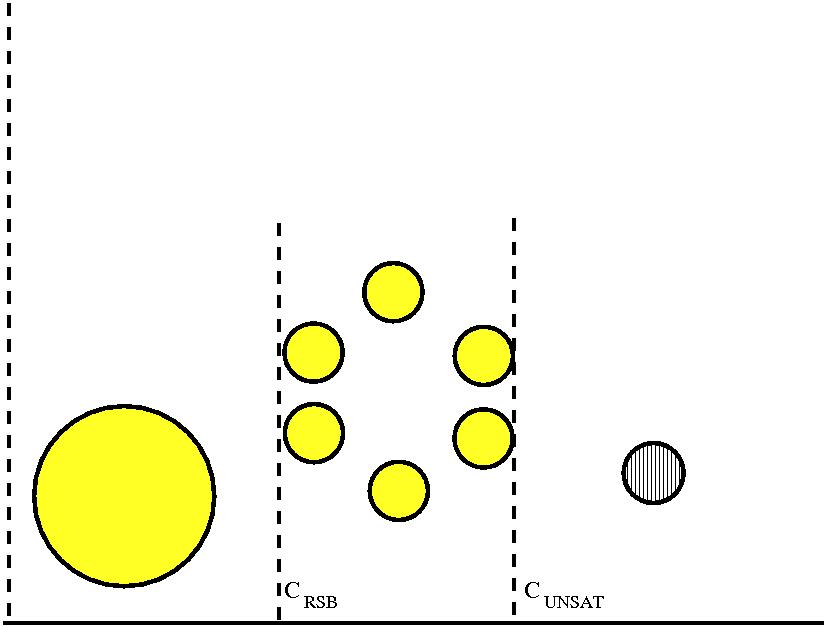}
\end{minipage}
\begin{minipage}[r]{6cm}
\includegraphics[height=4cm,width=6cm]{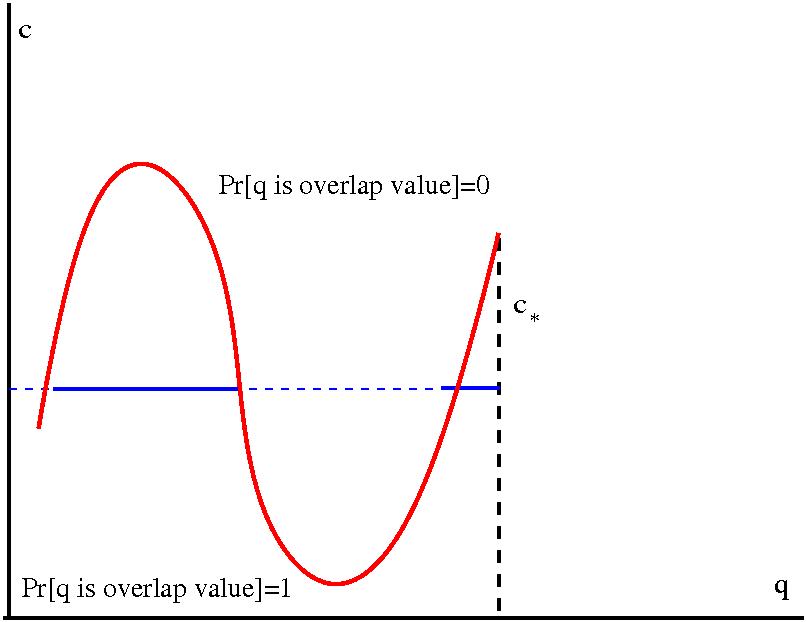}
\end{minipage}
\end{center}
\caption{ {\small{(a) Structure of solution space according to 1RSB
predictions.(b) Graphical description of the method used in
\cite{cond-mat/0504070/prl} to prove the discontinuity of support of
the  overlap distribution. }} }\label{fig:1-3}
\end{figure}

A great deal of insight in the complexity of random constraint
satisfaction problems has come from studying {\em phase transitions}
\cite{monasson:zecchina}. Concepts from Statistical Physics, such as
{\em first-order phase transitions}, {\em backbones}, or {\em
replica symmetry breaking} have helped to refine (and understand the
limitations of) the empirical observation that the ``hardest''
instances are located at the transition point. In some cases the
connection predicted by Statistical Physics can be made explicit in
purely combinatorial terms. For instance Monasson et al.
\cite{2+p:nature,2+p:rsa} have suggested that {\em first-order phase
transitions} are correlated with exponential complexity of
Davis-Putnam algorithms on random {\em unsatisfiable} instances at
the phase transition. This has been rigorously confirmed to a
certain extent \cite{2+p-sat,achlioptas-jcss,aimath04}. As for
instances in the satisfiable phase, much of the intuition on the
complexity of such instances comes again from Statistical Physics,
via the so-called {\em one step replica symmetry breaking} (1-RSB)
approach. The 1-RSB approach provides predictions on the geometric
structure of the set of satisfying assignments of a random formula
on $n$ variables. The set of such assignments can be naturally
viewed as a subgraph of the {\em hypercube of dimension $n$}, where
two satisfying assignments are {\em neighbors} if they only differ
in the value of one variable. Physics considerations imply that for
small values of the constraint density $c$ the set of satisfying
assignments forms a single cluster. The distribution of overlaps is
peaked around a certain constant value. The range of possible
overlaps (even those that are exponentially infrequent) is a
continuous interval. In the presence of 1-RSB, for constraint
density values higher than a critical value $c_{RSB}$ (smaller than
the unsatisfiability threshold $c_{UNSAT}$) the set of satisfying
assignments splits into several {\em clusters} such that: (i)
assignments in the same cluster all agree on a set of variables
having linear size. The distribution of overlaps of assignments in
the same cluster is still concentrated around a constant; (ii)
assignments in different clusters differ in $\Omega(n)$ variables;
(iii) the distribution of all overlaps has discontinuous support
(see Fig.~\ref{fig:1-3} (a); note that recent studies
\cite{krzakala07} suggest the existence of further phases below
$c_{RSB}$, omitted for simplicity from discussion and the figure).
The geometry of satisfying assignments outlined above has
implications for the complexity of heuristics such as local search,
algorithms such as belief propagation, or Davis-Putnam. The 1-RSB
approach provides (nonrigorous) values for the location of the phase
transition in random $k$-SAT \cite{threshold-cavity-rsa} that seems
to match the experimental evidence.  Algorithms that take advantage
of the geometry of solution space predicted by 1-RSB (e.g. the
celebrated {\em survey propagation} algorithm
\cite{surveyprop-rsa2005}) have greatly extended the range of
instances that can be solved in
practice.\\
 Rigorous results on the cluster structure of solutions of random
 CSP are emerging: M\'{e}zard et al. \cite{cond-mat/0504070/prl} have
developed an ingenious method for proving that the distribution of
overlap values of random $k$-SAT, with $k\geq 9$ indeed has
discontinuous support. Their approach is based on the following
concepts:

\begin{definition}
The {\em overlap} of two assignments $A$ and $B$ for a formula
$\Phi$ on $n$ variables, denoted by $overlap(A,B)$, is the fraction
of variables on which the two assignments agree (this is similar to
\cite{cond-mat/0504070/prl} and linearly related to the notion of
overlap from the statistical physics literature, where truth values
are modeled by $+1$ and $-1$, instead of 0/1). Formally
$overlap(A,B)=\frac{|\{i:A(x_{i})=B(x_{i})\}|}{n}. $
\end{definition}

The distribution of overlaps is, indeed, the original order
parameter that was originally used to study the phase transition in
random $k$-SAT \cite{monasson:zecchina}.

\begin{definition}
{\bf $q$-overlap-$k$-SAT}: Given a $k$-CNF formula $\Phi$ on $n$
variables, decide whether $\Phi$ has two satisfying assignments $A$
and $B$ such that $overlap(A,B)\in [q-1/\sqrt{n},q+1/\sqrt{n}]$
(following the suggestion in \cite{cond-mat/0504070/prl}, we will
use the function $1/\sqrt{n}$ for the the width of the possible
overlap around $q$; as discussed there, similar results are obtained
with any "reasonably large" function $b(n)=o(n)$). We will refer to
this event as {\em $A$ and $B$ have overlap approximately equal to
$q$}.
\end{definition}

For every value of $q$, the probability that a random $k$-SAT
formula has two assignments with overlap $\sim q$ is monotonically
decreasing with constraint density, and is empirically changing from
1 to 0 around a critical value $c_{k,q}$ of the constraint density.
If one can show that the function $W: q \goesto c_{k,q}$ is {\em
not} monotonic then {\em there exists a critical value $c_{*}$ such
that the horizontal line at $c_{*}$ will intersect the graph of
function $W$ at multiple points. Therefore (Figure~\ref{fig:1-3}
(b)) the distribution of overlaps in a random $k$-SAT formula of
constraint density $c_{*}$ has discontinuous support} (these results
were further extended, for $k$-SAT, $k\geq 9$, by Achlioptas and
Ricci-Tersenghi \cite{achlioptas-frozen}).

Our ultimate goal is to obtain an understanding of the underlying
reasons for the emergence of clustering in random CSP, with an
attempt at a precise classification. We investigate the nature of
overlap distributions of CSP under the random model defined and
investigated by Molloy \cite{molloy-stoc2002}. We cannot obtain a
complete classification (whether the results  in
\cite{cond-mat/0504070/prl,achlioptas-frozen} extend to random 3-SAT
is a more subtle problem; see \cite{maneva-sinclair07}). Instead, we
prove two partial results: Theorem~\ref{overlap-dichotomy-threshold}
shows that the first step of M\'{e}zard's approach can be applied to
all random CSP problems with a sharp threshold. In contrast, in
Theorem~\ref{replica-symmetry} we show that satisfying assignments
of random instances of 2-SAT in the satisfiable phase form a single
cluster, and can yield all possible values of the overlap. This
confirms the prediction \cite{monasson:zecchina} that the solutions
space of 2-SAT has a different nature, describes by the so-called
"replica symmetric" approach. The two results above are also
naturally related to results of Gopalan et al.
\cite{elitza-connectivity}. They proved a dichotomy theorem for the
complexity of deciding whether the set of satisfying assignments of
a CSP is connected (under the usual notion of adjacent assignments).
One ingredient of the result is a restriction (called {\em
tightness}) on the nature of constraints involved. Theorem 2
provides a  natural examples of CSP with tight constraints for which
there is evidence (the continuity of the overlap distribution) that
symmetry breaking does not take place. It also shows that, to be
really meaningful, the definition of adjacent assignments from
\cite{elitza-connectivity} should be somewhat modified.

\vspace{-2mm}
\section{Preliminaries}
\vspace{-2mm} Throughout the paper we will assume familiarity with
the general concepts of phase transitions in combinatorial problems
(see e.g. \cite{martin:monasson:zecchina}) and random structures.
One paper whose concepts and methods we use in detail (and we assume
greater familiarity with) is \cite{friedgut:k:sat}. Consider a
monotonically increasing problem $A=(A_{n})$ under the {\em constant
probability model} $\Gamma(n,p)$. For $\epsilon >0$ let
$p_{\epsilon}= p_{\epsilon}(n)$ define the canonical probability
such that $\Pr_{x \in \Gamma(n,p_{\epsilon}(n))}[x \in A]=
\epsilon$. The probability that a random sample $x$ satisfies
property $A$ (i.e. $x\in A$) is a monotonically increasing function
of $p$. Problem $A$ has a {\em sharp threshold} iff for every
$0<\epsilon < 1/2$, we have $lim_{n\goesto \infty}
\frac{p_{1-\epsilon}(n)- p_{\epsilon}(n)}{p_{1/2}(n)} = 0$. $A$ has
{\em a coarse threshold} if for some $\epsilon > 0$ it holds that
$\underline{\lim}_{n\goesto \infty} \frac{p_{1-\epsilon}(n)-
p_{\epsilon}(n)}{p_{1/2}(n)} > 0$. Related definitions can be given
for the other two models for generating random structures, the {\em
counting model} and {\em the multiset model}
\cite{bol:b:random-graphs}. Under reasonable conditions
\cite{bol:b:random-graphs} these models are equivalent, and we will
liberally switch between them. In particular, for satisfiability
problem $A$, and an instance $\Phi$ of $A$, $c_{A}(\Phi)$ will
denote its {\em constraint density}, the ratio between the number of
clauses and the number of variables of $\Phi$. To specify the random
model in this latter cases we have to specify the constraint density
as a function of $n$, the number of variables. We will use  $c_{A}$
to denote the value of the constraint density $c_{A}(\Phi)$ (in the
counting/multiset models) corresponding to taking $p=p_{1/2}$ in the
constant probability model. $c_{A}$ is a function on $n$ that is
believed to tend to a constant  as $n\goesto \infty$. However,
Friedgut's proof \cite{friedgut:k:sat} of a sharp threshold in
$k$-SAT (and our results) leave this issue open.

\begin{definition}\label{model} Let ${\cal D} = \{0,1,\ldots, t-1\}$, $t\geq 2$
be a fixed set. Consider the set of all $2^{t^{k}}-1$ potential
nonempty binary constraints on $k$ variables $X_{1}, \ldots, X_{k}$.
We fix a set of constraints ${\cal C}$ and define the random model
$CSP({\cal C})$.  A random formula from $CSP_{n,p}({\cal C})$ is
specified by the following procedure: (i) $n$ is the number of
variables; (ii) for each $k$-tuple of {\em ordered} distinct
variables $(x_{1}, \ldots, x_{k})$ and each $C\in {\cal C}$ add
constraint $C(x_{1}, \ldots, x_{k})$ independently with probability
$p$. We will write $SAT({\cal C})$ instead of $CSP({\cal C})$ for
{\em boolean} constraint satisfaction problems (i.e. $t=2$).
\end{definition}

\begin{definition}\label{overlap-model} Let ${\cal D} = \{0,1,\ldots, t-1\}$, $t\geq 2$ be
a fixed set. Let $q$ be a real number in the range [0,1]. The
problem $q$-overlap-$CSP({\cal C})$ is the decision problem
specified as follows: (i) The input is an instance $\Phi$ of
$CSP_{n,p}({\cal C})$; (ii) The decision problem is whether $\Phi$
has two satisfying assignments $A,B$ such that $ overlap(A,B)\in
[q-1/\sqrt{n}, q+1/\sqrt{n}]$ (following \cite{cond-mat/0506053}, we
will informally refer to the property as {\em ``$\Phi$ is
$q$-satisfiable''}). The random model for $q$-overlap-$CSP({\cal
C})$ is simply the one for $CSP_{n,p}({\cal C})$. We will refer to
this class of problems as {\em fixed-overlap CSP}.
\end{definition}

The notion of adjacent satisfying assignments used in
\cite{achlioptas-frozen}, while adequate for random $k$-SAT, is not
suited for other random CSP. For instance, it is impossible to flip
exactly one bit in a satisfying assignment of an instance of
1-in-$k$ SAT \cite{1-in-k} and still obtain a satisfying assignment
(except for the  case when that variable does not appear in the
formula). Thus we will use the following setup: let $f(n)=o(n)$ be a
suitably large function; we will assume that $\lim f(n)/\log n =
\infty$. Two satisfying assignments that differ on at most $f(n)$
variables will be called adjacent. A {\em cluster} is a connected
component of the set of satisfying assignments.
\section{Results}
In this section we study the sharpness of the threshold for random
generalized constraint satisfaction problem defined by Molloy
\cite{molloy-stoc2002}.  Creignou and Daud\'{e}
\cite{creignou-daude-threshold} (and independently the author of this
paper \cite{istrate-sharp}) have characterized the boolean CSP
problems $SAT({\cal C})$ with a sharp threshold:

\begin{definition} A set of constraints ${\cal C}$ is {\em interesting}
if there exist constraints  $C_{0},C_{1}\in {\cal C}$ with
$C_{0}(\overline{0})=C_{1}(\overline{1})=0$, where
$\overline{0},\overline{1}$ are the "all zeros" ("all ones")
assignments. Constraint $C_{2}$ is {\em an implicate of $C_{1}$} iff
every satisfying assignment for $C_{1}$ satisfies $C_{2}$.  A
boolean
 constraint $C$ {\em strongly depends on a literal} if it has an unit
 clause as an implicate. A boolean constraint $C$ {\em strongly
 depends on a 2-XOR relation} if $\exists i,j\in \{1,\ldots,k\}$ such
 that constraint ``$x_{i}\neq x_{j}$'' is an implicate of $C$.
\end{definition}

\begin{proposition}\label{dichotomy-threshold} \cite{creignou-daude-threshold,istrate-sharp}
Consider a generalized satisfiability problem $SAT({\cal C})$ with
${\cal C}$ interesting. (i) If some constraint in ${\cal C}$
strongly depends on one literal then $SAT({\cal C})$ has a coarse
threshold; (ii) If some constraint in ${\cal C}$ strongly depends on
a 2XOR-relation then $SAT({\cal C})$ has a coarse threshold; (iii)
In all other cases $SAT({\cal C})$ has a sharp threshold.
\end{proposition}

Mora et. al \cite{cond-mat/0506053} proved that all problems
$q$-overlap-$k$-SAT, $k\geq 2$ have a sharp threshold.  We extend
this result by showing that for all CSP with a sharp threshold,
their fixed-overlap versions also have a sharp threshold:

\begin{theorem}\label{overlap-dichotomy-threshold}
Consider a generalized satisfiability problem $SAT({\cal C})$ such
that (i) ${\cal C}$ is interesting (ii) No constraint in ${\cal C}$
strongly depends on a literal; (iii) No constraint in ${\cal C}$
strongly depends on a 2XOR-relation. Then for all values $q\in
(0,1]$ the problem $q$-overlap-$SAT({\cal C})$ has a sharp
threshold.
\end{theorem}

The previous result does not yet rigorously prove the existence of
curve $W$ since it does not prove  fact that the phase transition in
the $q$-overlap versions happens at some constant constraint density
$c_{q}$ .

Given the previous result, how can a problem $SAT({\cal C})$ have an
overlap distribution with continuous support ? Obviously, the second
step of the approach in \cite{cond-mat/0504070/prl} must fail. This
happens when {\em the location $c_{q}$ of the transition for the
$q$-overlap version of $SAT({\cal C})$ is a monotonic function of
the overlap $q$}. The next result shows gives a natural problem for
which this is indeed the case:

\begin{theorem}\label{replica-symmetry} The following are true:
\begin{itemize}
\item[(i)] Let $c<1$. Then with probability $1-o(1)$ the satisfying assignments of a
random instance of $2$-SAT of constraint density $c$ form a single cluster.

\item[(ii)] Also, let $q\in (0,1]$. Let $c< 1$. Then with probability $1-o(1)$ a
random instance of $2$-SAT of constraint density $c$ is $q$-satisfiable.
\end{itemize}
\end{theorem}

\section{Proof of Theorem~\ref{overlap-dichotomy-threshold}}

Before presenting the proof, let us remark that for boolean
constraints, the hypothesis of the
Theorem~\ref{overlap-dichotomy-threshold} implies that the set of
constraints ${\cal C}$ is {\em well-behaved}. That
is\cite{molloy-stoc2002}, every formula whose hypergraph is
tree-like or unicyclic is satisfiable. This is, for instance, an
easy consequence of conditions (D0),(D1), Theorem 4.1 in
\cite{creignou-daude-threshold}. Also, since ${\cal C}$ is
interesting there exist constraints $\Gamma_{0}, \Gamma_{1}\in {\cal
C}$ such that $\Gamma_{0}(x_{1}, \ldots, x_{k})\models
\overline{x_{1}}\OR \ldots \OR \overline{x_{k}}$ and
$\Gamma_{1}(x_{1}, \ldots, x_{k})\models x_{1}\OR \ldots \OR x_{k}$.

We will employ the Friedgut-Bourgain criterion for the existence of
a sharp threshold of a monotonic property $A$. Note that any problem
$q$-overlap-SAT(${\cal C}$) is indeed monotone, since adding clauses
can only reduce the set of satisfying assignments, in particular
decreasing the probability of $q$-satisfiability. The starting point
of all applications of the Friedgut-Bourgain criterion is noting
that if a monotone property $A$ has a coarse threshold then there
exists $0< \epsilon < 1/2$, $p^{*}=p^{*}(n)\in [p_{1-\epsilon},
p_{\epsilon}]$ and $C>0$ such that $p \cdot
\frac{d\mu_{p}(A)}{dp}|_{p = p^{*}(n)} < C$. Bourgain and Friedgut
have shown that the following holds:

\begin{proposition} \label{sufficient-sharp}
Suppose $p= o(1)$ is such that $p \cdot \frac{d\mu_{p}(A)}{dp}|_{p =
p^{*}(n)} < C$. Then there is $\delta = \delta(C)>0$ such that
either $ \mu_{p}(x\in \{0,1\}^{n} | \mbox{ }x\mbox{ contains
}x^{\prime}\in A\mbox{ of size }|x^{\prime}|\leq 10C\} > \delta$, or
there exists $x^{\prime}\not \in A$ of size $|x^{\prime}| \leq 10C$
such that $ \mu_{p}(x\in A| x \supset x^{\prime})>\mu_{p}(A)+
\delta$.
\end{proposition}

(in fact, in \cite{friedgut:k:sat} the proposition is stated
assuming for convenience that $p=p_{1/2}$, but this is not needed.
We give here the general statement). We will need, in fact, an
enhancement to the Bourgain-Friedgut result that was given by
Friedgut in \cite{friedgut-survey-sharp}: For a finite set of words
$W$ define {\em the filter generated by $W$}, $F(W)$ as
$F(W)=\{x\mbox{ }|\mbox{ }(\exists y \in W)\mbox{ with }x\supseteq
y\}$. Friedgut noted (\cite{friedgut-survey-sharp}, remarks on pages
5-6 of that paper) that the set $W$ of ``booster'' sets $x^{\prime}$
in the second conditions satisfies $\mu_{p}(F(W))=\Omega(1)$.

Consider now a set of constraints ${\cal C}$ satisfying the
conditions the Theorem, and let
$A=\overline{\mbox{q-overlap-SAT(}{\cal C})}$. Applying
Proposition~\ref{sufficient-sharp} enhanced by the previous
observation, and taking into account the fact that the number of
isomorphism types of formulas of size at most $10C$ is finite, we
infer that we can assume that formula $x^{\prime}$ in the second
condition appears with probability $\Omega(1)$ as a subformula in a
random formula in $\mbox{q-overlap-SAT}_{p}({\cal C})$. Furthermore,
instead on conditioning on the presence of $x^{\prime}$ as a subset
of $x$ one can, instead, add it. Finally, note that for random
constraint satisfaction problems, because of the invariance of such
problems under variable renaming, one only needs to add a random
copy of $x^{\prime}$. Putting all these observations together, the
following version of Proposition~\ref{sufficient-sharp} holds:

\begin{proposition} \label{sufficient-sharp-3}
Suppose  $p= o(1)$ is such that $p \cdot \frac{d\mu_{p}(A)}{dp}|_{p
= p^{*}(n)} < C$. Then there is $\delta = \delta(C)>0$ such that
either
\begin{equation}\label{first-condition-sharp-3}
\mu_{p}(x\in \{0,1\}^{n} | \mbox{ }x\mbox{ contains }x^{\prime}\in A\mbox{ of size }|x^{\prime}|\leq 10C\} > \delta
\end{equation}

or there exists $F\not \in A$ of size $|F| \leq 10C$,
such that
\begin{itemize}
\item Formula $F$ appears with probability $\Omega(1)$ as a subformula
in a random formula in $CSP_{p}({\cal C})$.
\item If $\Xi$ denotes the formula obtained by creating a copy of
$x^{\prime}$ on a random set of variables, then

\begin{equation}
\label{second-condition-sharp-3}
 \mu_{p}(x\cup \Xi \in A)>\mu_{p}(A)+ \delta.
\end{equation}
\end{itemize}
\end{proposition}

To show that random $q\mbox{-overlap-SAT(}{\cal C})$ has a sharp
threshold, we will reason by contradiction. Assuming this is not the
case, one needs to prove that the two conditions in
Proposition~\ref{sufficient-sharp-3} do not hold.

Suppose, indeed, that condition~(\ref{first-condition-sharp-3}) was
true. That is, with positive probability it is true that a random
formula $\Phi\in CSP({\cal C})$ contains some subformula
$\Phi^{\prime} \in \overline{\mbox{q-overlap-SAT(}{\cal C})}$ of
size at most $10C$. With high probability all subformulas of a
random formula $\Phi$ of size at most $10C$ are either tree-like or
unicyclic.  But because the set of constraints ${\cal C}$ is
well-behaved (this is the point where the hypothesis on the
constraint set ${\cal C}$ is used), all formulas in $CSP({\cal C})$
that are tree-like or unicyclic are satisfiable. Since the formula
contains a finite number of variables, one can set the other
variables not appearing in $\Phi$ in a way that will create two
satisfying assignments with overlap approximately $q$. Therefore the
first condition in Proposition~\ref{sufficient-sharp-3} cannot be
true.

Assume, now, that condition~(\ref{second-condition-sharp-3}) is
true. That is, there exists $F\in q\mbox{-overlap-SAT(}{\cal C})$, a
formula of size at most $10C$, such that adding $F$ to a random
formula $\Phi\in CSP_{p}({\cal C})$ diminishes the probability that
the resulting formula has two assignments of overlap $\simeq q$ by
at least a constant $\delta$. As discussed, we assume that $F$
occurs with probability $\Omega(1)$ in a random formula in
$CSP_{p}({\cal C})$. Therefore $F$ is tree-like or unicyclic.

\begin{definition}
A {\em unit clause} is a constraint (not necessarily part of the
constraint set ${\cal C}$) specified by a condition $X=\delta$, with
$X$ being a variable and $\delta \in \{0,1\}$.
\end{definition}

\begin{lemma}
If $F$ satisfies condition ~(\ref{second-condition-sharp-3}) then
there exists another formula $G$ that is specified by a finite
conjunction of unit clauses $
G\equiv (X_{1}=\delta_{1})\AND \ldots \AND (X_{p}=\delta_{p})$,
that also satisfies condition~(\ref{second-condition-sharp-3}).
\end{lemma}

\beginproof
Formula $F$ appears with constant probability in a random $CSP({\cal
C})$ formula with probability $p$ and has constant size. Therefore
$F$ is either tree-like or unicyclic. The result follows easily by
replacing $F$ with formula $G$ consisting of the conjunction of unit
constraints corresponding to a satisfying assignment of $F$. Indeed,
$G$ is tighter than $F$, so adding a random copy of $G$ instead of a
random copy of $F$ can only increase the probability that the
resulting formula is unsatisfiable.
\endproof\qed

\vspace{-5mm} The key to refuting
condition~(\ref{second-condition-sharp-3}) is to show that, if it
did hold then, for every monotonically increasing function $f(n)$
that tends to infinity, we could also increase the probability of
unsatisfiability by a positive constant if, instead of conditioning
on $x$ containing a copy of $F$,  we add $f(n)$ random constraints
from set ${\cal C}$. We first prove:

\begin{lemma}\label{connecting-sharp-2}

Let $0<\tau < 1$ be a constant and let $p$ be such that
$\mu_{p}(q-overlap-SAT({\cal C}))\geq \tau$. Assume that $r\geq 1$
and that $g_{1}, g_{2}, \ldots g_{r}$ are elements of $\{0,1\}$ such
that, when $(X_{1}, X_{2}, \ldots, X_{r})$ is a random $r$-tuple of
different variables \vspace{-2mm}
\begin{equation}\label {hypothesis-connecting-2}
Pr(\Phi\mbox{ has sat. assign. } A,B\mbox{ of overlap } \simeq
q\mbox{ with } X_{1}=g_{1}, \ldots, X_{r}=g_{r}) \leq \frac{\tau}{2}
.
\end{equation}

Then there exists constant $m\geq 1$ (that only depends on
$k,r,\tau$) such that, if $\eta$ denotes a formula from $CSP({\cal
C})$ obtained by adding, for each $x\in\{0,1\}$, $m\cdot r\cdot
2^{k^{r}}$ random copies of $\Gamma_x$, then \vspace{-2mm}
\begin{equation} \label{third-condition-sharp-2}
Pr(\Phi \cup \eta \in q\mbox{-overlap-SAT(}{\cal C})) \leq \frac{\tau}{2}
\end{equation}
\end{lemma}
\vspace{-2mm}

\beginproof

For $i\in \{1,\ldots, r\}$ define $A_{i}$ to be the event that the
formula $\Phi$ has a pair of satisfying assignments of overlap
$\simeq q$ with $X_{1}=g_{1}, \ldots, X_{i}=g_{i}$. Also define
$A_{0}$ to be the event that $\Phi\in \mbox{q-overlap-SAT(}{\cal
C})$. The hypothesis translates as the fact that both inequalities
$Pr(A_{0})\geq \tau$ and $Pr(A_{r}) \leq \frac{\tau}{2}$ are true.
Therefore $ Pr(A_{r}|A_{0})= \frac{Pr(A_{r}\AND
A_{0})}{Pr(A_{0})}\leq \frac{\tau/2}{\tau} =\frac{1}{2}$.  Since
$\overline{A_{r-1}}\implies \overline{A_{r}}$  we have
\begin{equation}\label{initial-ineq}
\mu_{r} := Pr[\overline{A_{r}}|A_{0}]=
Pr[\overline{A_{r-1}}|A_{0}]+Pr[\overline{A_{r}}|A_{r-1}\AND
A_{0}]\cdot Pr[A_{r-1}|A_{0}]\geq \frac{1}{2}
\end{equation}

But $Pr[\overline{A_{r}}|A_{r-1}\AND
A_{0}]=Pr[\overline{A_{r}}|A_{r-1}]$ is the fraction of variables in
formula $\Phi \AND (X_{1}=g_{1})\AND \ldots \AND (X_{r-1}=g_{r-1})$
that have to receive values different from $g_{r}$ in order for the
resulting formula to still have two satisfying assignments of
overlap $\sim q$; let $C_{r}$ be the set of such variables.  If
instead of the last unit constraint we add a random copy of
constraint $\Gamma_{g_{r}}$, the resulting formula is in
$\overline{\mbox{q-overlap-SAT(}{\cal C})}$ when all the variables
appearing in the new constraint are in the set $C_{r}$. Denoting
$\lambda_{r}= Pr[\overline{A_{r}}|A_{r-1}]$, the probability of this
last event happening is $\lambda_{r}^{k}/(1-o(1))$ (we choose a
$k$-tuple of distinct variables from a set of density
$\lambda_{r})$; Thus the probability that the new formula is in
$\overline{\mbox{q-overlap-SAT(}{\cal C})}$ is at least $\nu_{r}:=
Pr[A_{r-1}|\overline{A_{0}}]+\frac{\lambda_{r}^{k}}{1-o(1)}\cdot
Pr[\overline{A_{r-1}}|\overline{A_{0}}]$. Applying Jensen's
inequality to the convex function $f(x)=x^{k}$ and using
inequality~(\ref{initial-ineq}), we infer
\begin{eqnarray*}
\frac{1}{2^k} & \leq & \mu_{r}^{k}=
(Pr[A_{r-1}|\overline{A_{0}}]\cdot
1+Pr[A_{r}|\overline{A_{r-1}}]\cdot
Pr[\overline{A_{r-1}}|\overline{A_{0}}])^{k} \leq \\ & \leq &
Pr[A_{r-1}|\overline{A_{0}}]\cdot
1^{k}+Pr[A_{r}|\overline{A_{r-1}}]^{k}\cdot
Pr[\overline{A_{r-1}}|\overline{A_{0}}] = \\ & = &
(Pr[A_{r-1}|\overline{A_{0}}] +\lambda_{r}^{k}\cdot
Pr[\overline{A_{r-1}}|\overline{A_{0}}])=\nu_{r}\cdot (1+o(1)).
\end{eqnarray*}

Thus $\nu_{r}\geq \frac{1}{2^k}\cdot (1-o(1))$. The conclusion of
this long argument is that adding one random copy of
$\Gamma_{b_{r}}$ instead of the $r$-th constraint lowers the
probability of membership to $q\mbox{-overlap-SAT(}{\cal C})$ to no
less than $\frac{1}{2^k}\cdot (1-o(1))$. Adding the copy of the
constraint {\em before} the first $r-1$ unit constraints and
repeating the argument recursively implies the fact that, if instead
of adding the $r$ unit constraints to $\Phi$ we add $r$ random
copies of $\Gamma_{b_{1}}, \ldots, \Gamma_{b_{r}}$ that the
resulting formula belongs to $\overline{\mbox{q-overlap-SAT(}{\cal
C})}$, given that $\Phi\in q\mbox{-overlap-SAT(}{\cal C})$, is at
least $\gamma_{r} = \frac{1}{2^{k^{r}}(1-o(1))}$. Since the values
$b_{1}, \ldots, b_{r}$ can repeat themselves, the same is true if we
add $r$ random copies of $\Gamma_x$ for {\em every}  $x$.

Suppose now that we add $r\cdot m\cdot 2^{k^{r}}$ copies of each
$\Gamma_x$ (that is, we repeat the random experiment $m\cdot
2^{k^{r}}$ times, for some integer $m\geq 1$). The probability that
none of the experiments will make the resulting formula
unsatisfiable is at most $(1-\gamma_{r})^{m\cdot 2^{k^{r}}}$. For
some constant $m$ this is going to be at most $1-\frac{\tau}{2}$.
This means that $Pr(\Phi \cup \eta \mbox{ is satisfiable}) \leq
\frac{\tau}{2}$.\endproof\qed \vspace{-2mm}
 We can refute condition~(\ref{third-condition-sharp-2}) directly,
thus obtaining a contradiction. To do so, we employ the following
result (Lemma 3.1 in \cite{achlioptas:friedgut:kcol}):

\begin{lemma}\label{achlioptas:friedgut} For a monotone property
\footnote{Achlioptas and Friedgut assume $A$ to be a monotone {\em graph}
property, but this fact is not used anywhere in their proof.}  $A$ let
$\mu(p)= Pr[G\in \Gamma(n,p)\mbox{ has property }A]$, and let
$\mu^{+}(p,M)=Pr[G_{1}\cup G_{2}\mbox{ }|\mbox{ }G_{1}\in \Gamma(n,p),
G_{2}\in \Gamma(n,M)\mbox{ has property }A]$.

Let $A=A(n)\subseteq \{0,1\}^{n}$ be a monotone property and $M=M(n)$
such that $M= o(\sqrt{np})$. Then $|\mu(p)-\mu^{+}(p,M)|= o(1)$.
\end{lemma}

We obtain a contradiction in the following way: consider a random
formula $\eta$ with $f(n)$ clauses, for some $f(n)\goesto \infty$.
It is easy to show that the probability that $\eta$ contains, for
some $x$, less than $r\cdot m\cdot 2^{k^{r}}$ copies of  $\Gamma_x$
(with $r,m$ as in Lemma 2) is $o(1)$. So adding $\eta$ (instead of
the random formula in Lemma~\ref{connecting-sharp-2}) decreases the
probability of  $q$-satisfiability by at least $\delta-o(1)$. But
this contradicts the conclusion of Lemma~\ref{achlioptas:friedgut}.
\qed \vspace{-5mm}
\section{Proof of Theorem~\ref{replica-symmetry}}

We will use the well-known graph-theoretic interpretation of 2-CNF
formulas, that associates to a given formula $\Phi$ on $n$ variables
a directed graph $G_{\Phi}$ with $2n$ vertices $\{x_{1}, \ldots,
x_{n},\overline{x_{1}}, \ldots, \overline{x_{n}}\}$, and for every
clause $C=\alpha \OR \beta$ of $\Phi$ it adds directed edges
$\overline{\alpha}\goesto \beta$ and $\overline{\beta}\goesto
\alpha$ to $G_{\Phi}$.  We will need a number of  results from
\cite{franco-pure-2sat} concerning the structure of graph $G_{\Phi}$
when $\Phi$ is a random formula of constraint density $c<1$.

\begin{definition}
A {\em cycle} is a set $l_{1}\goesto l_{2}, l_{2}\goesto l_{3},
\ldots, l_{s}\goesto l_{1}$ of directed edges. Two cycles $C_{1},
C_{2}$ are {\em overlapping} if they share at least an edge. Two
cycles $C_{1}, C_{2}$ are {\em connected by a path} if there exist
vertices $x\in C_{1}, y\in C_{2}$ and a path (possibly of length zero,
i.e. $x=y$) from $x$ to $y$.
\end{definition}

\begin{lemma}\label{claim1}
Let $t=t(n)$ such that $1=o(t)$. Let $\Phi$ be a random 2-CNF formula
of constraint density $c<1$ and $G_{\Phi}$ be its associated digraph.
With probability $1-o(1)$ the following are true: (i) $G_{\Phi}$
contains no cycles connected by a path.  (ii) $G_{\Phi}$ contains no
overlapping cycles. (ii) the sum of all the cycle lengths is less
than $t$.
\end{lemma}

To these results we add the following claim (whose proof is similar
to that of Claim~\ref{claim1} (i) from \cite{franco-pure-2sat}):
With probability $1-o(1)$ no literal implies literals in two
different cycles.

We can thus divide the literals of the formula into four classes:
(i) those that are on a cycle. (ii) those that are not on a cycle,
but {\em imply} a literal on a cycle. (iii) those that are not on a
cycle, but are {\em implied} by a literal on a cycle. (iv) those
that are not on a cycle and neither imply nor are implied by a
literal on a cycle.

\begin{definition}
A literal $x$ is {\em bad} if there exists $y$ such that
$x\stackrel{*}{\goesto} y$, $x \stackrel{*}{\goesto} \overline{
y}$.
\end{definition}

We first claim that there is a function $h(n)=o(n)$ such that with
probability $1-o(1)$ the number of bad literals is at most $h(n)$.
Indeed, all bad literals can only be set to false in any satisfying
assignment of the formula. This means that a bad literal belongs to
the {\em spine} of the formula \cite{scaling:window:2sat}. But a
standard argument (see e.g. \cite{aimath04}) shows that the size of
the spine is $o(n)$.

 Bad literals (and their negations) are assigned fixed
values in all satisfying assignments. This property guarantees that
such literals do not influence the value of the overlap between any
two satisfying assignments. Let ${\cal B}$  be the set of such
literals.\\
 {\bf Theorem 2(i):}  Let $A$ and $B$ be two satisfying assignments of a
formula $\Phi$, such that $d(A,B)> \log n$ (i.e. $A$ and $B$ are not
adjacent). We will prove the following result:

\begin{lemma}
  There exists a satisfying assignment $C$ such that $d(A,C)= O(\log n)$ and $d(C,B)<d(A,B)$. That is, $C$ is adjacent to $A$ and closer to $B$ than $A$.
 \end{lemma}

 An iterative application of the lemma proves the Theorem 2(i).

\noindent{\bf Proof:} Let $x$ be a variable such that $A(x)\neq
B(x)$ and $x$ is {\em implication minimal} with this property.
  In other words if $y\neq x$ and $y\tcimplies x$ then $A(y)=B(y)$.

\noindent{\bf Case 1:} $A(x)= 0$ and $B(x)= 1$. Then $B(z)=1$ for
all $z$ such that $x\tcimplies z$. Define
 the assignment $C$ by $C(z)=1$ if $x\tcimplies z$, $C(z)=A(z)$ otherwise.
 It is clear that $d(C,B)<d(A,B)$, since $C$ coincides with $B$ on all bits whose value changes.
 To show that $C$ is a satisfying assignment, suppose $C$ did not satisfy
 some clause $W=(\alpha \OR \beta)$. Then one of the following is true.
 \begin{itemize}
 \item[(1): ] both $\alpha$ and $\beta$ are negations of literals implied by $x$.
 This leads to a contradiction, since it would imply that $B$ does not satisfy clause
 $\alpha \OR \beta$ either.
 \item[(2): ] one of them (say $\alpha$) is the negation of a literal implied by $x$. Since
 $x\tcimplies \overline{\alpha}$ and $\overline{\alpha} \rightarrow \beta$, it follows that $C(\beta)=1$,
 so $C$ satisfies clause $W$.
 \item[(3): ] none of them is the negation of a literal implied by $x$. Then $C(\alpha)=A(\alpha)$ and
 $C(\beta)=A(\beta)$, a contradiction, since $A$ satisfies clause $W$.
  \end{itemize}

\noindent{\bf Case 2:} $A(x)= 1$ and $B(x)= 0$. Then $B(z)=0$ for
all $z$ such that $z\tcimplies x$. Define
 the assignment $C$ by $C(z)=0$ if $z\tcimplies x$, $C(z)=A(z)$ otherwise. It is clear that $d(C,B)<d(A,B)$, since $C$ coincides with $B$ on all the bits that change value,
 one of which is $x$. To show that $C$ is a satisfying assignment, suppose $C$ did not satisfy
 some clause $\alpha \OR \beta$. Then one of the following cases must hold
 \begin{itemize}
 \item[(1): ] both $\alpha$ and $\beta$ are literals that imply $x$. This leads to a contradiction,
 since this would mean that $B$ with respect to satisfying clause $\alpha \OR \beta$.
 \item[(2): ] one of them (say $\alpha$) implies $x$. Since $\overline{\beta} \tcimplies \alpha$, it
 follows that $\overline{\beta} \tcimplies x$, therefore $\beta$ is assigned the value TRUE by $C$, a
 contradiction.
 \item[(3): ] none of $\alpha,\beta$ implies $x$. Then $C$ and $A$ coincide
 with respect to the values they give to $\alpha,\beta$, a contradiction, since $A$ satisfies clause $W$.
  \end{itemize}

{\bf Theorem 2(ii):} We directly construct two satisfying
assignments $A$ and $B$ of overlap $qn\pminus \sqrt{n}$.  We will
work with a directed weighted graph $G_{2}$ obtained from $G_{\Phi}$
by contracting every cycle to a node and assigning this node a
weight equal to twice the size of the contracted cycle. $G_{2}$ is
well-defined when cycles in $G_{\Phi}$ do not intersect, an event
that happens (cf. Claim~\ref{claim1}) with probability $1-o(1)$. All
literals on a cycle of $G_{\Phi}$ need, of course, to be given the
same value in any satisfying assignment. Since we have contracted
all cycles in $G_{\Phi}$, $G_{2}$ is a directed acyclic graph.  The
set of nodes corresponding to bad literals is downward closed,
because if $x\goesto y$ and $y$ is bad then $x$ is bad.
Correspondingly, the set of nodes corresponding to negations of a
bad literal is upward closed.

We begin by defining a set $S$ of nodes of $G_{2}$ that will
ultimately contain half of the nodes in $G_{2}$. Nodes not chosen in
$S$ will be referred to as {\em eliminated}). In parallel we build a
partial assignment by assigning those literals corresponding to
eliminated nodes the unique values that are consistent with the
satisfiability of the formula. Set $S$ is recursively specified as
follows: (i) start by defining $V$ to be the set of all nodes in
$G_{2}$ (ii) add all nodes of of indegree $0$ in $V$ to $S$ and
eliminate all nodes of outdegree 0. Set $V$ to be the set of
remaining nodes (not added to $S$ or eliminated).
 (ii) continue this process as long as $V\neq \emptyset$.

It is easy to see that the set of literals corresponding to nodes in
$S$ contains, for every variable $x$, exactly one of $x$ and
$\overline{x}$. Indeed, one cannot add both $x$ and $\overline{x}$ to
$S$ in one step, otherwise the pure literal implying both would be
bad. But then, when adding one of them we immediately eliminate the
other one. On the other hand, we only eliminate a literal when its
opposite has been retained in $S$.

The first assignment, $A$ simply corresponds to setting all literals
corresponding to nodes in  $S$ to TRUE. We define the second
assignment iteratively by the following process: (i) in Stage $1$
choose a node of indegree zero, assign its associated variable the
value FALSE and eliminate the node from $S$. If the eliminated node
corresponds to a cycle in $G_{2}$ all variables in the cycle are set
to FALSE.  (ii) when a remaining node becomes of indegree zero as a
result of eliminations, it is labeled by the value of the stage that
led to this happening (nodes that originally had indegree zero are
labelled 0).  (iii) the literal chosen to set to FALSE is among
those with a smallest stage number. (iv) continue the process until
the number of variables assigned FALSE is in the interval
$[qn-\sqrt{n}, qn+\sqrt{n}]$. This is possibly if the sum of all
cycle lengths in the formula graph of $\Phi$ is $o(\sqrt{n})$, which
happens (cf. Lemma~\ref{claim1} ) with probability $1-o(1)$.  (v)
The remaining literals in $S$ are set to TRUE.

Because bad literals are assigned identical values in both $A$ and
$B$ it is easy to see that $overlap(A,B)\in [qn-\sqrt{n},
qn+\sqrt{n}]$. We complete the proof of
Theorem~\ref{replica-symmetry} by:

\begin{lemma}
$A$ and $B$ are satisfying assignments for $\Phi$.
\end{lemma}

{\bf Proof:} Suppose there exists a clause $C= (\overline{x}\OR
y)\equiv (x\goesto y)$ of $\Phi$ that is not satisfied by $A$. Then
$x$ is given a TRUE value and $y$ is given a FALSE value. Thus
either $\overline{x}$ is a bad literal, or $x$ is in $S$. Also,
either $y$ is a bad literal or $\overline{y}$ is in $S$. Suppose $y$
were a bad literal. Then, since $x\goesto y$, $x$ is also bad. But
this contradicts the two possible alternatives ($\overline{x}$ is a
bad literal or $x$ is in $S$). Suppose now $\overline{y}$ is in $S$.
Then $C\equiv (\overline{y}\goesto \overline{x})$. Therefore, either
$\overline{x}\in S$ or $\overline{x}$ is among the literals (bad
literals and their negations) eliminated before defining $S$. The
first alternative leads to a contradiction with the two possible
alternatives ($\overline{x}$ is a bad literal or $x$ is in $S$), so
it must be that $\overline{x}$ is a bad literal. But then
$\overline{y}$ is also bad, contradicting the assumption that
$\overline{y}$ is in $S$.

A similar argument shows that $B$ is a satisfying assignment. Indeed, suppose
 there existed a clause $C= (\overline{x}\OR y)\equiv (x\goesto y)$ of
 $\Phi$ not satisfied by $B$. Then $B(x)=TRUE$, $B(y)=FALSE$. The
 choices compatible with this setup are: (i) $x$ is in $S$ and
 $B(x)= TRUE$, or $\overline{x}$ is bad.  (ii) $y$ is bad, or $y$ is
 in $S$ and $B(y)=FALSE$, or $\overline{y}\in S$ and
 $B(\overline{y})=TRUE$, i.e. $B(y)=FALSE$. First, if $y$ were bad then so would be $x$, contradicting all
possible choices in (i). If $x,y$ were both in $S$, with $y$
assigned FALSE, since by construction of $B$ the set of literals in
$S$ is downward closed under implication it follows that $x$ would
also be assigned FALSE, a contradiction. The other other possibility
is that $\overline{x}$ is bad. But since $\overline{y} \goesto
\overline x$ that would mean that $\overline{y}$ is bad, a
contradiction with the assumption that $y\in S$. Finally, assume
$\overline{y}$ is in $S$ and is assigned TRUE. Since $\overline{y}
\goesto \overline{x}$ either $\overline{x}\in S$ or $x$ is a bad
literal. In the first case, since the set of literals assigned to
TRUE is upward closed under implication it would mean that
$\overline{x}$ is assigned TRUE by $B$, i.e. $x$ is assigned FALSE,
a contradiction. Suppose now that $x$ is bad.  Then $B(x)=0$, a
contradiction. \qed

\vspace{-10mm} {\small
\newcommand{\etalchar}[1]{$^{#1}$}
}


\end{document}